# METHODOLOGY FOR INTELLIGENT INJECTION POINT LOCATION BASED ON GEOMETRIC ALGORITHMS AND DISCRETE TOPOLOGIES FOR VIRTUAL DIGITAL TWIN ENVIRONMENTS

Jorge Manuel Mercado-Colmenero[1], Abelardo Torres – Alba[2] y Cristina Martin-Doñate[1*]

1 Department of Engineering Graphics Design and Projects. University of Jaen. Spain

2 INGDISIG Jaén Research Group, Campus Las Lagunillas s/n. A3 Building, Jaén 23071, Spain

*Correspondence: cdonate@ujaen.es

Campus Las Lagunillas, s/n. Building A3-210 23071 Jaen (Spain) Phone: +34 953212821, Fax: +34 953212334

**ABSTRACT:**

*Implementing intelligent design models can revolutionize the use of digital twins, which are crucial in product design by incorporating intelligent algorithms. This perspective is especially important for the design of injection-molded plastic parts, a complex process that often requires expert knowledge and costly simulation software not available to all companies. This article presents an innovative methodology for locating injection points in injection-molded parts using intelligent models with geometric algorithms for discrete topologies. The first algorithm calculates the center of mass of the discrete model based on the center of mass of each triangular facet in the system, ensuring uniform molten plastic distribution during mold cavity filling. Two sub-algorithms intelligently evaluate the geometry and optimal injection point location. The first sub-algorithm generates a geometric matrix based on a two-dimensional nodal quadrature adapted to the part's bounding box. The second sub-algorithm projects the nodal matrix and associated circular areas orthogonally on the part's surface along the demolding direction. The optimal injection point location is determined by minimizing the distance to the center of mass from the first algorithm's result. This novel methodology has been validated through rheological simulations in six case studies with complex geometries. The results demonstrate uniform and homogeneous molten plastic distribution with minimal pressure loss during the filling phase. Importantly, this methodology does not require expert intervention, reducing time and costs associated with manual injection mold feed system design. It is also adaptable to various design environments and virtual twin systems, not tied to specific CAD software. The validated results surpass the state of the art, offering an agile alternative for digital twin applications in new product design environments, reducing dependence on experts, facilitating designer training, and ultimately cutting costs*

RESUMEN:

La implementación de modelos de diseño inteligente revoluciona el uso de gemelos digitales en el diseño de productos, incorporando algoritmos inteligentes. Esto es especialmente relevante para piezas plásticas moldeadas por inyección, un proceso complejo que a menudo requiere conocimientos expertos y software costoso de simulación no accesible para todas las empresas. Presentamos una metodología para localizar el punto de inyección en piezas moldeadas por inyección utilizando modelos inteligentes con algoritmos geométricos para topologías discretas. El primer algoritmo calcula el centro de masas del modelo discreto basado en el centro de masas de cada faceta triangular en el sistema, garantizando una distribución uniforme de plástico fundido durante el llenado del molde. Dos subalgoritmos evalúan la geometría y la ubicación óptima del punto de inyección. El primero genera una matriz geométrica basada en una cuadratura nodal bidimensional adaptada a la pieza. El segundo proyecta la matriz nodal y áreas circulares asociadas de forma ortogonal en la superficie de la pieza a lo largo de la dirección de desmoldeo. La ubicación óptima del punto de inyección se determina minimizando la distancia al centro de masas según los resultados del primer algoritmo. Esta metodología se ha validado mediante simulaciones reológicas en seis casos de estudio con geometrías complejas, demostrando una distribución uniforme del plástico fundido con una pérdida de presión mínima durante el llenado. Lo más importante es que esta metodología no requiere expertos, reduciendo tiempo y costos en el diseño del sistema de alimentación del molde de inyección. Además, es adaptable a diferentes entornos de diseño y sistemas de gemelos virtuales, sin necesidad de un software CAD específico. Los resultados validados superan el estado del arte, ofreciendo una opción ágil para gemelos digitales en el diseño de nuevos productos, reduciendo la dependencia de expertos, facilitando la capacitación de diseñadores y, en última instancia, reduciendo costos.

Palabras clave: : Moldeo por Inyección, Diseño de Productos, Diseño Industrial, Algoritmos Geométricos, Gemelo Digital

**FUNDING**

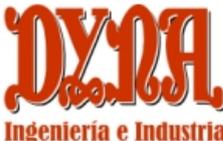

| ARTICULO DE INVESTIGACIÓN | METHODOLOGY FOR INTELLIGENT INJECTION POINT LOCATION BASED ON GEOMETRIC ALGORITHMS AND DISCRETE TOPOLOGIES FOR VIRTUAL DIGITAL TWIN ENVIRONMENTS | Industrial Technology Process Technology |
|---|---|---|
|  | Jorge Manuel Mercado-Colmenero, Abelardo Torres – Alba, Cristina Martin-Doñate |  |

The research has been funded by the University of Jaén through the Research Plan 2022-2023-ACTION1a POAI 2022-2023: TIC-159, and by the scientific society INGEGRAF through the authors' award for research presented in the field of industrial design and manufacturing in injection moulding (2023).

## 1.- INTRODUCCION

Digital twins are dynamic virtual models connected to physical systems throughout the product lifecycle, with the aim of exchanging data between them in an automated way [1]. These digital twins can be used at all stages of design [2], enabling the use of different graphical models adapted to each phase. Currently, digital product twins focus primarily on optimising the functional and technological features of components through virtual simulations [3]. Unfortunately, simulation software may not be available to the company or there may be a lack of highly experienced designers to solve difficult cases. Intelligent design models characterised by autonomy and complex thinking greatly assist in the process of creating virtual digital geometries. This is especially applicable to the design of plastic parts manufactured by injection moulding, a complex process that requires in-depth knowledge of the component geometry, as well as rheological, thermal and mechanical aspects inherent to the manufacturing process [4-6]. In the design of plastic injection moulds, the proper choice of the location for the material gate is essential to ensure balanced flow during injection [7-8]. In the current mould design process, it is common for designers to choose potential injection gate locations on the surface of the CAD model, using heuristic methods to identify the optimal location [9]. However, this approach has multiple drawbacks, since the efficiency of the final design is highly dependent on the experience of the designer, who usually considers a limited set of locations. Due to the lack of efficient algorithms and the problems encountered when trying to obtain the gate location in 3D models, researchers and scientists have focused their efforts on developing more effective methods for determining the gate location in injection moulded plastic parts.

Two mathematical approaches are generally used to determine the location of the injection point on an injection moulded part. The first approach is based on limiting the choice zone to a specific region of the surface by setting the upper and lower limits on the coordinate axes. However, this method is only applicable for simple cases on horizontal or vertical surfaces of the part. The second approach consists of manual selection of the injection point location in more complex areas followed by the application of optimisation algorithms to find the optimal location. Although this second approach is more widespread than the first, in practice, the localisation is limited by the predefined space, which restricts the search to a limited number of pre-selected locations. Hsu et al. [10] propose a novel methodology based on sequential input optimisation using a parallel efficient global optimisation algorithm. Zhay et al. [11] present a general methodology called flow resistance search scheme to determine the optimal gate location in the plastic injection moulding process. This approach develops a search direction based on flow resistance by exploiting the physical characteristics of the mould filling. Hamdy et al. [12] carry out a full three-dimensional analysis for a mould with a cuboid-shaped cavity of various thicknesses. The results show that the location of the inlet normal to the minimum cavity thickness minimises the time required for complete solidification of the product. Liu et al. [13] introduce a hybrid optimisation algorithm to find a set of inlet locations in liquid composite moulding. Shen et al. [14] develop an optimisation strategy based on a genetic algorithm. The objective function considers the injection pressure at the end of mould filling. Some authors make use of the Voronoi diagram for injection point location, among them Fu et al. [15] present a concurrent optimisation algorithm that aims to improve the structural geometry and gate location in the design of multi-material parts for the injection moulding process. This algorithm introduces a new technique based on Voronoi diagrams to simulate the formation of weld lines, thus allowing the identification of interfaces between different materials. The work presented by Wang et al. [16] uses an iterative method and the centroidal Voronoi diagram to find the ideal gate locations in the resin transfer moulding process, with the aim of minimising the mould filling time. Unfortunately, in the set of algorithms presented, the solution search space remains a challenge when the selected region is either arbitrary in shape or difficult to describe by a parametric function.

With the aim of improving and optimising the current process of plastic part analysis and injection mould design, the article presents a new intelligent methodology based on the use of geometric algorithms for discrete topologies developed by the authors that allows the location and sizing of the injection gate in a plastic part without the need for manual interaction by the user. This innovative methodology for mould cavity gate design overcomes the state of the art by reducing reliance on expert designers, facilitating the training of new designers and ultimately lowering the associated costs. Furthermore, the methodology is not tied to any particular CAD software, making it flexible and easily adaptable to various design environments or for virtual digital twin systems for plastic components or injection moulds.

## 2.- MATERIALS AND METHODS
### 2.1. GEOMETRIC ALGORITHMS FOR AUTOMATED LOCATION OF THE INJECTION POINT ON THE PLASTIC PART

In order to optimise the design phases of the elements and systems involved in the injection moulding process, the methodology proposed in this article proposes the use of the STL (Stereolithographic) geometric format as a reference for the geometry of the plastic part. This format consists of a discrete geometric model composed of vertices and triangular surfaces or facets that describe the external surface of the part. As shown in Figure 1, the STL format provides a surface mesh that is independent of the CAD software where the plastic part geometry is generated. From this discrete geometric model, a set of intelligent geometric algorithms is defined to establish the optimal location of the injection point of the plastic part. This allows the molten plastic front to develop uniformly, homogeneously, with the least loss of pressure and temperature along the injection mould cavity during its filling phase, thus avoiding the possible formation of defects on the surface of the plastic part.

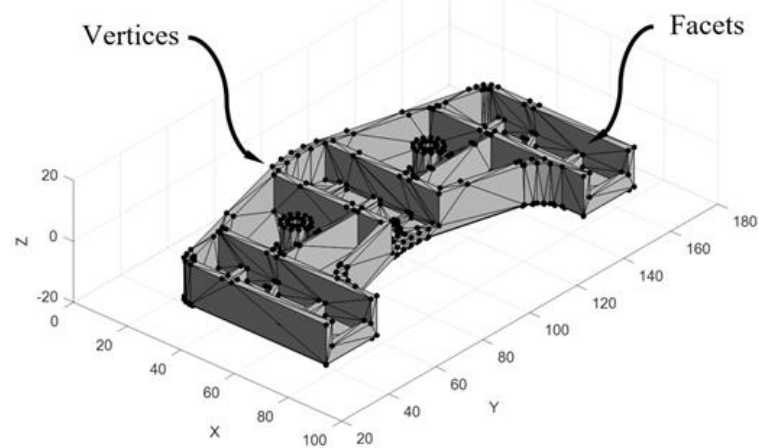

*Fig. 1. Definition of the geometrical described in stl format*

This section focuses on the geometric algorithms developed throughout the proposed research work to determine the optimal location of the injection point of a plastic part. According to the filling phase of a plastic part, the molten plastic front introduces the injection mould cavity through the gate geometry or injection point. Subsequently, it advances along the injection mould cavity and gradually occupies the entire volume, thus completing the filling or injection phase. However, during this filling process, the molten plastic front experiences variations in its pressure and temperature field. These fluctuations are attributed to or influenced by the trajectory of the molten plastic front along the inside of the injection mould cavity. Equation (1) [17] determines that the pressure loss experienced by the molten plastic front inside the injection mould cavity is directly related to its path length. Similarly, the heat transfer Fourier equation, Equation (2) [17], establishes a direct relationship between the temperature of the molten plastic front and its path length within the injection mould cavity. Consequently, as the path length of the molten plastic front increases, both the pressure drop inside the injection mould cavity and the temperature difference of the molten plastic front increase simultaneously, see Equation (1) and Equation (2).

$$\Delta P = \frac{12 \cdot \mu \cdot L \cdot \bar{v}}{H^2} \quad (1)$$

$$\frac{dT}{dt} = \alpha \frac{d^2 T}{d^2 x} \quad (2)$$

Where ΔP [MPa] represents the pressure drop across the injection mould cavity, μ [Pa-s] represents the viscosity of the molten thermoplastic material, L [mm] represents the path length of the molten plastic front across the injection mould cavity, $\bar{v}$ [mm/s] represents the average velocity of the molten plastic front, H [mm] represents the thickness of the plastic part and α [mm$^2$/s] represents the thermal diffusivity of the thermoplastic material.

Therefore, it is established that an optimal design of the injection point must minimise the distance travelled by the molten plastic front along the injection mould cavity during its filling phase. This hypothesis determines the criteria for defining the automated geometric algorithms proposed in this article. Moreover, the optimal location of the injection point influences the improvement of the uniformity of the rheological parameters of the molten plastic front during the mould cavity filling phase. Thus, it is important to note that the injection point is located as close as possible to the centre of mass of the plastic part. This minimises the distance travelled by the molten plastic front along the inside of the mould cavity during its filling phase. Thus, the first algorithm developed by the authors establishes, in an automated manner, the centre of mass of the discrete geometry of the plastic part using the principle of superposition. This principle is based on the individual calculation of the centre of mass of each triangular facet of the discrete geometry of the plastic part and then using this information to obtain the resulting centre of mass of the entire plastic part. The superposition principle ensures that the centre of mass is calculated accurately, even in complex or irregularly shaped geometries. The centre of mass serves as an initial reference point for subsequent algorithms that locate the final location of the injection point, ensuring that it is optimally located for uniform and homogeneous development of the molten plastic front and ensuring that the minimum number of defects are generated on the surface of the final plastic part. As shown in Figure 2, the discrete geometry of the plastic part is defined as $C^f_{mesh}$, and is itself composed of a set of vertices and triangular facets. Each triangular facet has an associated area and centre of mass, defined as $A^f$ and $C^f_{CM}$ respectively. The definition of these geometric parameters is established as shown in Equation (3) and Equation (4).

$$A^f = \tfrac{1}{2} \cdot \left\| (V_2^f - V_1^f) \cdot (V_3^f - V_1^f) \right\| \quad (3)$$

$$C^f_{CM} = \tfrac{1}{3} \cdot (V_3^f + V_2^f + V_1^f) \quad (4)$$

Where $V^f_3$, $V^f_2$, $V^f_1$ represent the vertices of each triangular facet (see Figure 2). By using Equation (3) and Equation (4), the centre of mass of the entire plastic part, $C_{CM}$, can be calculated by the principle of superposition, as shown in Equation (5).

$$\forall C^f_{mesh} \exists C_{CM} : C_{CM} = \frac{\sum C^f_{CM} \cdot A^f}{\sum A^f} \quad (5)$$

The principle of superposition states that the centre of mass of a discrete geometry can be determined from the centres of mass and area associated with each division. Using Equation (3), Equation (4) and Equation (5), the centre of mass of the plastic part can be calculated efficiently and accurately.

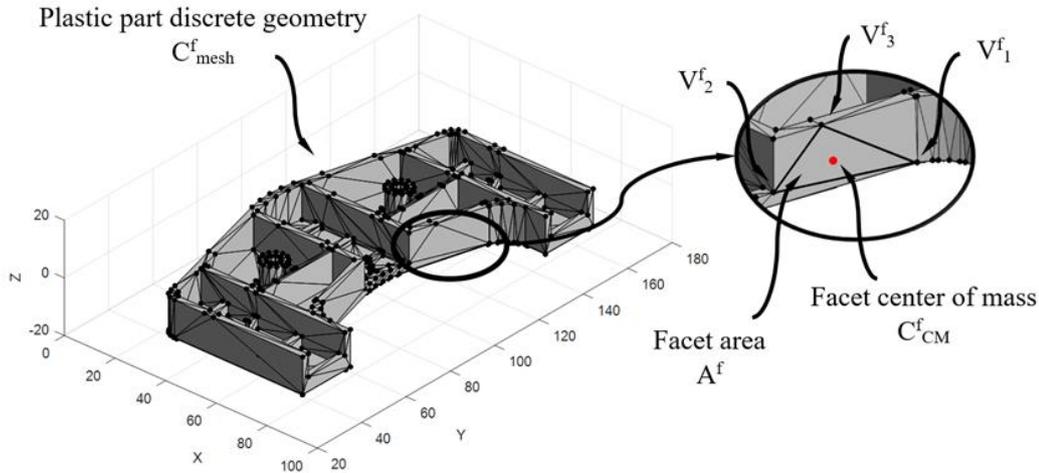

*Fig. 2. Definition of the centre of mass of the discrete geometry of the plastic part*

Next, the second geometric algorithm focuses on intelligently locating the Cartesian coordinates of the point on the surface of the plastic part that is closest to its centre of mass and that also meets the design criteria of the injection gate. For this purpose, the geometric algorithm used in this research work is responsible for generating a 2D nodal quadrature, called $\lambda_q$, comprising a set of nodes defined as $\delta_q$. The accuracy of the 2D nodal quadrature is equal to the minimum dimensional or geometric detail of the plastic part. Each of these nodes $\delta_q$ is assigned an associated circular area $A(\delta_q)$, established in Equation (6), and whose diameter coincides with the diameter established in Equation (8) for the sizing of the geometry of the injection point, assuming that the geometry of the inlet section is circular.

$$\forall \delta_q \exists A(\delta_q) : A(\delta_q) = \begin{bmatrix} A(\delta_q)_x = R_{gate} \cdot Cos(\theta) + \delta_{qx} \\ A(\delta_q)_y = R_{gate} \cdot Sin(\theta) + \delta_{qy} \end{bmatrix} \quad (6)$$

Where $R_{gate}$ [mm] represents the radius of the circular cross-section of the inlet. In addition, in order to dimension the radius of the circular inlet section, Equation (7) is used to determine the average velocity of the molten plastic front along the filling of the injection mould cavity [17].

$$\bar{v} = \sqrt{\frac{5 \cdot (T_{melt} - T_{wall}) \cdot \kappa}{3 \cdot \mu_{opt}}} \quad (7)$$

Where $T_{melt}$ [ºC] represents the temperature of the molten plastic front, $T_{wall}$ [ºC] represents the temperature of the mould surface, κ [W/m-ºC] represents the thermal conductivity of the thermoplastic material and $\mu_{opt}$ [Pa-s] represents the viscosity magnitude of the thermoplastic material to obtain an optimal filling phase. It should be noted that the magnitudes of these parameters are set as recommendations by the manufacturer of the thermoplastic material. Then, from the expression for the average velocity of the molten plastic front, see Equation (7), and with the equation that determines the shear rate of a non-Newtonian fluid, $\dot{\gamma}_{opt}$ [1/s], the equation that establishes the optimum radius of the circular inlet is defined, as shown in Equation (8).

$$\dot{\gamma}_{opt} = \frac{(3+1/n) \cdot \bar{v}}{R_{gate}} \longrightarrow R_{gate} = \frac{(3+1/n) \cdot \sqrt{\frac{5 \cdot (T_{melt} - T_{wall}) \cdot \kappa}{3 \cdot \mu_{opt}}}}{\dot{\gamma}_{opt}} \quad (8)$$

It should be noted that, in the case of sizing a rectangular cross-section for the inlet geometry, the hydraulic radius of a rectangular cross-section must be equated with the radius of the circular inlet geometry obtained from Equation (8). Next, after sizing the geometry associated with the injection point, the proposed methodology continues from the nodal quadrature $\lambda_q$. The points of the nodal quadrature $\delta_q$ together with the points of the associated circular area $A(\delta_q)$ are projected according to the demoulding direction $D_d$ onto the discrete geometry of the plastic part $C_{mesh}$ (see Figure 3) in order to avoid locating the injection point in unacceptable areas of the plastic part surface. The set of points projected on the mesh $C_{mesh}$ are evaluated by selecting those points $\delta_q$ and the corresponding points associated with the circular area $A(\delta_q)$, which intersect the mesh $C_{mesh}$ as a whole.

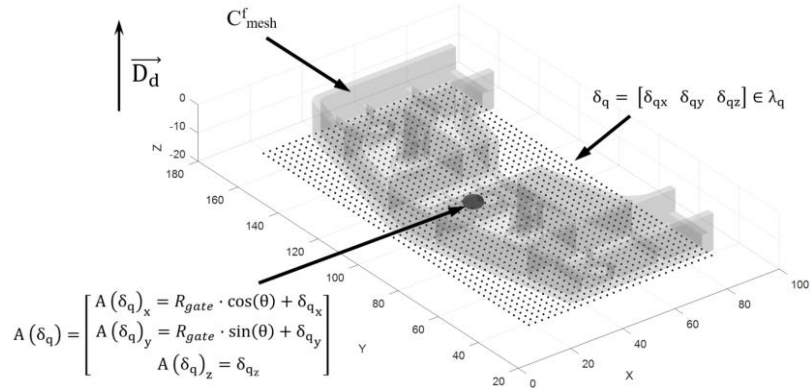

**Fig. 3.** *Representation of the points $\delta_q$ of the 2D nodal quadrature $\lambda_q$ and example of the area $A(\delta_q)$ associated to a point $\delta_q$*

The definition of the radius of the circular inlet $R_{gate}$ as a function of the diameter of the area $A(\delta_q)$ determines the safety distance of the geometric algorithm. Thus, in order to meet the design requirements of the injection mould, the magnitude of the circular inlet radius $R_{gate}$ must be smaller than the thickness of the plastic part, thus ensuring that the entire cross-section of the feed channel is contained in $A(\delta_q)$. Finally, the injection point $C_{pointfill}$ is defined from the points $\delta_q$ and the associated circular areas $A(\delta_q)$, which intersect the discrete geometry $C_{mesh}$, selecting the point $\delta_q$ that is located closest to the centre of mass of the plastic part $C_{CM}$ (see Figure 4). In this way, the drag and the pressure and temperature drop of the molten plastic front are reduced, ensuring a balanced flow with an even distribution of the injection pressure along the injection mould cavity during its filling phase. In case the part to be manufactured has aesthetic requirements, the algorithm places the injection point $C_{point\ fill}$ at the point on the parting line [20] closest to the resulting centre of mass of the plastic part, see Figure 5.

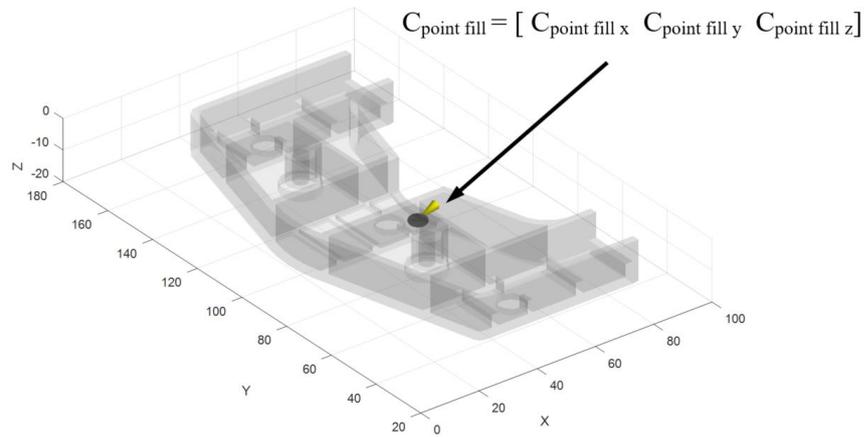

*Fig. 4.* Result of the geometric algorithms for the intelligent location of the inlet in plastic parts

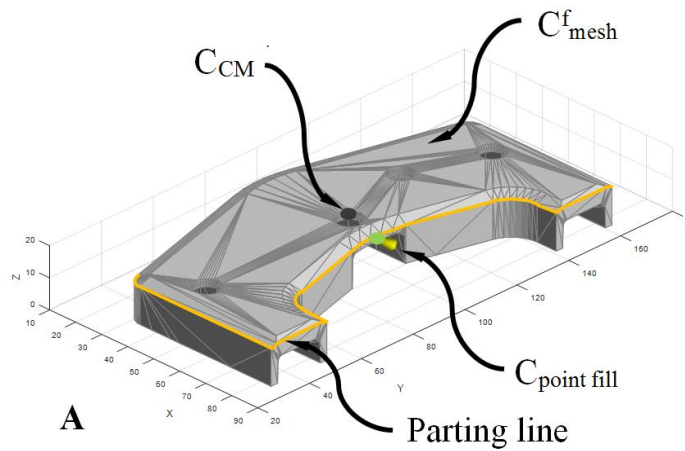

*Fig. 5.* Results of the intelligent geometric algorithms including aesthetic requirements.

## 3. RESULTS OF INTELLIGENT GEOMETRIC ALGORITHMS AND COMPARISON WITH NUMERICAL CAE SIMULATIONS

In this section, the results obtained from the intelligent geometric algorithms proposed in this research article are shown. Likewise, these analytical results are compared with the numerical results obtained by means of rheological simulations carried out using CAE software. The numerical results determine, from the CAD model of the plastic part to be manufactured, the optimal location of the injection point. The application of the intelligent geometric algorithms has been developed in Matlab R2017 software [18]. This implies that the geometric algorithms designed for the determination of the centre of mass of the plastic parts and the final location of the injection point have been developed in this computer software, taking advantage of its own programming language and the routines and functions included in it. On the other hand, the numerical simulations, used to compare and validate the results of the proposed methodology, have been carried out with the commercial software Autodesk Moldflow Adviser [19]. In order to evaluate the accuracy and reliability of the proposed methodology, six case studies covering different geometries and thermoplastic materials are analysed.

Table 1 shows the rheological and thermal parameters of the thermoplastic materials used for each of the case studies, as well as the resulting sizing result for their circular inlet radius.

| Case study | Material | n | $T_{melt}$ [ºC] | $T_{wall}$ [ºC] | $\dot{\gamma}_{opt}$ [1/s] | $\mu_{opt}$ [Pa·s] | κ [W/m·ºC] | $R_{gate}$ [mm] |
|---|---|---|---|---|---|---|---|---|
| 1 | PP | 0.2718 | 230 | 50 | 10,000 | 9.88 | 0.15 | 1.4 |
| 2 | PP | 0.2718 | 230 | 50 | 10,000 | 9.88 | 0.15 | 1.4 |
| 3 | ABS | 0.2354 | 230 | 50 | 5,000 | 30.93 | 0.18 | 1.9 |
| 4 | ABS | 0.2354 | 230 | 50 | 5,000 | 30.96 | 0.18 | 1.9 |
| 5 | PC | 0.1869 | 305 | 95 | 8,000 | 42.85 | 0.24 | 1.5 |
| 6 | PC | 0.1869 | 305 | 95 | 8,000 | 42.85 | 0.24 | 1.5 |

*Tabla 1. Parameters of the thermoplastic materials used and dimensioning of the circular inlet*

On the other hand, Figure 6, Figure 7, Figure 8, Figure 9, Figure 10 and Figure 11 show both the results generated by the intelligent algorithms developed in this article and the numerical results obtained from the rheological simulations carried out. It should be noted that the accuracy of the STL models for the definition of the discrete geometry of the plastic parts analysed has a deviation tolerance of 0.27 mm and an angular tolerance of 30º.

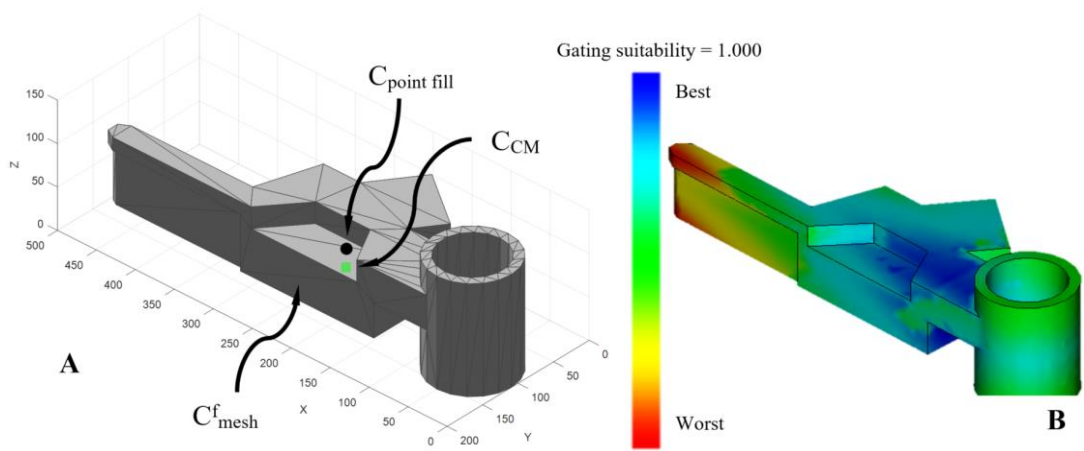

Fig. 6. (A) Result of the geometric algorithms and (B) Numerical simulation results for the location of the gate

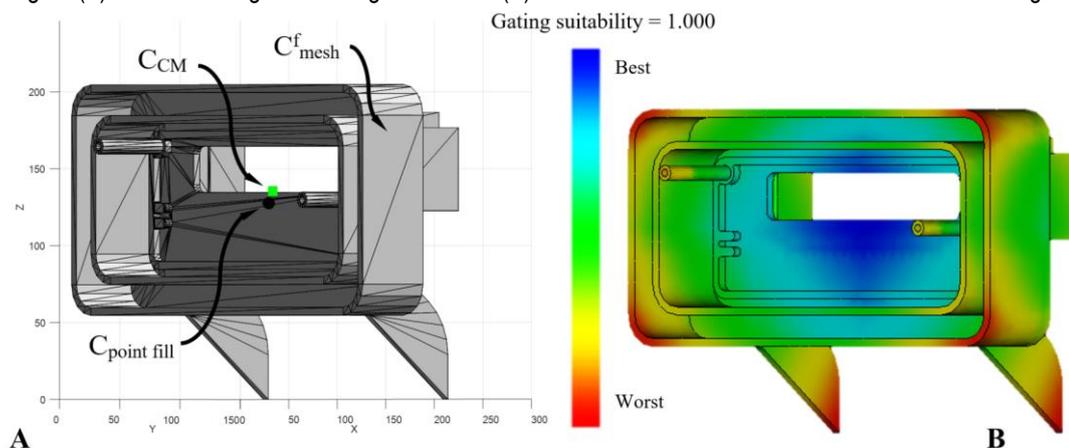

Fig. 7. (A) Result of the geometric algorithms and (B) Numerical simulation results for the location of the gate

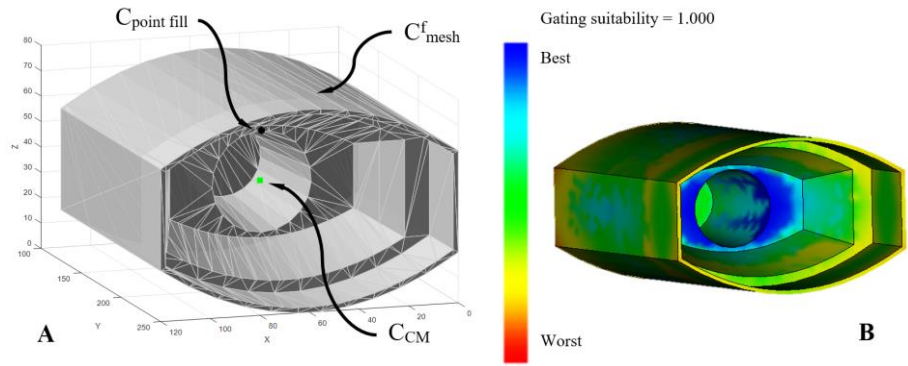

Fig. 8. (A) Result of the geometric algorithms and (B) Numerical simulation results for the location of the gate

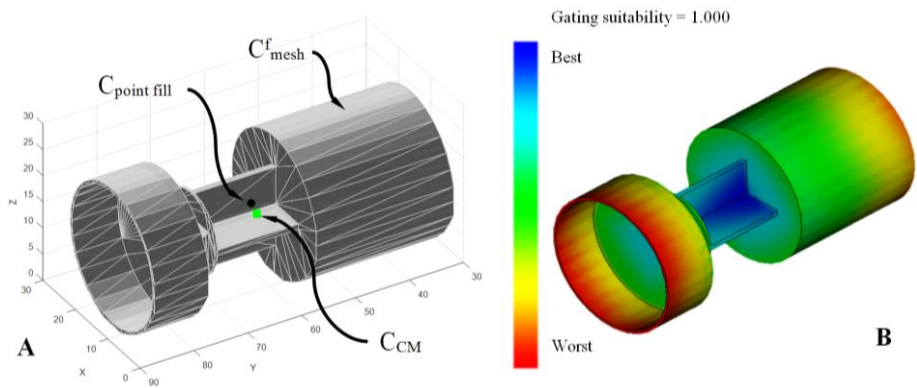

Fig. 9. (A) Result of the geometric algorithms and (B) Numerical simulation results for the location of the gate

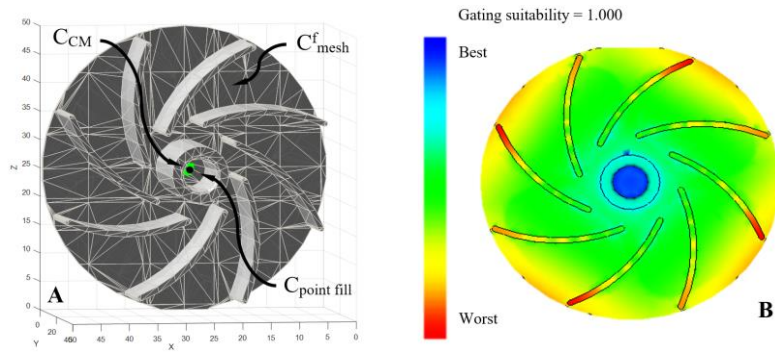

Fig. 10. (A) Result of the geometric algorithms and (B) Numerical simulation results for the location of the gate

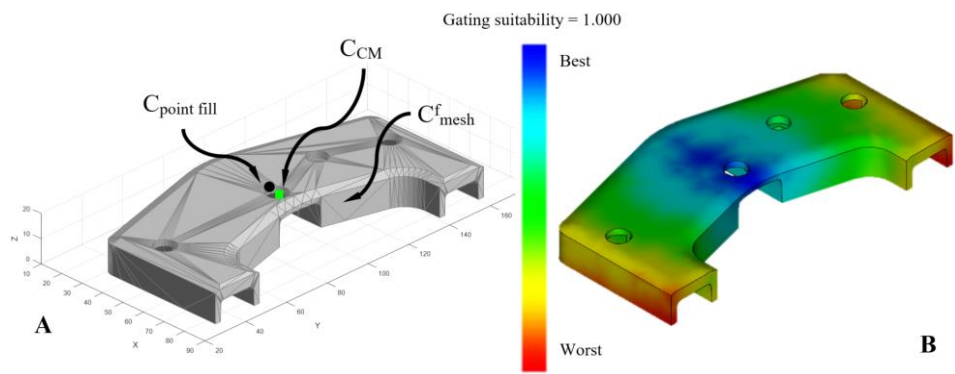

Fig. 11. (A) Result of the geometric algorithms and (B) Numerical simulation results for the location of the gate

| Case study | Intelligent algorithms [s] | Numerical simulations [s] |
|---|---|---|
| 1 | 13.8 | 149 |
| 2 | 13.4 | 157 |
| 3 | 13.2 | 164 |
| 4 | 6.2 | 162 |
| 5 | 9.5 | 156 |
| 6 | 14.6 | 163 |

*Tabla 2. Comparison of the computational demand obtained for both methodologies*

In order to quantitatively evaluate the improvement and computational efficiency of the intelligent algorithms developed with respect to the conventional process of selecting the injection point by means of numerical simulations, Table 2 shows the time elapsed for the definition of the injection point for both methodologies. For this purpose, an MSI computer with an Intel (R) Core (TM) i-77700HQ CPU @ 2.80 GHz has been used.

The results of the methodology presented in the paper reveal a complete agreement between the input location provided by the algorithm and that obtained through the simulations. These findings attest to the reliability and accuracy of the innovative methodology, which offers a fast and efficient alternative to the conventional manual injection mould feed system design process. As evidenced by the results in Table 2, the methodology has the advantage of being significantly faster and less computationally demanding compared to numerical simulations. The comparison demonstrates that this methodology can determine the inlet location in areas on the surface of the plastic part ensuring a uniform and homogeneous flow of the molten plastic, with minimal pressure drop during the filling phase of the injection mould cavity.

## 4. CONCLUSIONS

This paper presents a methodology based on intelligent geometric algorithms to determine the location of the injection point in the plastic injection moulding process. Two innovative geometric algorithms have been developed for this purpose. The first algorithm focuses on calculating the centre of mass of the discrete model from the sum of moments of the triangular facets that make up the discrete global system, using a superposition principle. The second algorithm performs an automated scan of the geometry in search of the optimal location for the injection inlet. The proposed methodology has been validated by comparing the results generated by the automated algorithm with those obtained through numerical simulations. For this purpose, five case studies with different geometries have been examined. The results show an excellent agreement between the input location obtained by the intelligent algorithm and the one determined by the simulations. These results corroborate the reliability and accuracy of the proposed methodology, which offers an agile and efficient alternative to the traditional manual design process of the injection mould feeding system. In this line, it has been possible to determine the location of the gate in areas on the surface of the plastic part that guarantee a uniform and homogeneous flow of the molten plastic, with a minimum pressure drop during the filling phase of the injection mould cavity. In addition, this methodology has the advantage of being considerably faster and requiring fewer computational resources compared to numerical simulations. The presented methodology introduces an innovative approach to injection mould injection point design by incorporating knowledge in the determination of the optimal injection point location. By taking advantage of intelligent design algorithms capable of analysing the discrete geometric model of the plastic part, this methodology provides the basis for the creation of an automated tool that eliminates the need for expert intervention in the determination of the injection point and reduces both product development time and costs associated with the conventional manual design process. Furthermore, this methodology is not tied to any specific CAD software, making it flexible and easily adaptable to a variety of manufacturing processes. In summary, the methodology presented in the article is a valuable tool for the design and development stages of plastic parts and can be implemented in smart digital twin applications in the field of product design.

## REFERENCIAS

| | METHODOLOGY FOR INTELLIGENT INJECTION POINT LOCATION BASED ON GEOMETRIC ALGORITHMS AND DISCRETE TOPOLOGIES FOR VIRTUAL DIGITAL TWIN ENVIRONMENTS | Industrial Technology |
|---|---|---|
| ARTICULO DE INVESTIGACIÓN | Jorge Manuel Mercado-Colmenero, Abelardo Torres – Alba, Cristina Martin-Doñate | Process Technology |